\documentclass[prb,showpacs,twocolumn,reprint,amsmath,amssymb]{revtex4-1}
\usepackage{amsmath}
\usepackage{graphicx}
\usepackage{color}
\usepackage{multirow}
\begin{document}
\title{Half-metallic Dirac cone in zigzag-graphene-nanoribbon/graphene}
\author {M. X. Chen}
\email{chen59@uwm.edu}
\affiliation{Department of Physics, University of Wisconsin, Milwaukee, Wisconsin 53211, USA}
\author {M. Weinert}
\affiliation{Department of Physics, University of Wisconsin, Milwaukee, Wisconsin 53211, USA}

\date{\today}

\begin{abstract}
The Dirac electrons of graphene, an intrinsic zero gap semiconductor,
uniquely carry spin and pseudospin that give rise to many fascinating
electronic and transport properties.  While isolated zigzag graphene
nanoribbons are antiferromagnetic semiconductors, we show by means of
first-principles and tight-binding calculations that zigzag graphene
nanoribbons supported on graphene are half-metallic as a result of spin-
and pseudospin-symmetry breaking.  In particular, half-metallic Dirac
cones are formed at K (K') near the Fermi level.  The present results
demonstrate that the unique combination of spin and pseudospin in zigzag
graphene nanoribbons may be used to manipulate the electronic properties
of graphene, and may have practical implications for
potential graphene-based nanoelectronic applications.
\end{abstract}

% 71.20.-b Band structure
% 73.22.Pr, electronic structure/condensed matter/graphene, 
% 73.20.-r, Electron states at surfaces and interfaces
% 75.75.-c Magnetic properties of nanostructures

\maketitle

\section{Introduction}
The discovery of graphene has inspired intense interest in graphene-based nanostructures 
and their derivatives such as van der Waals (vdW) heterostructures.\cite{novoselov_2005,geim_van_2013}
One fascinating feature of graphene is the combination of real spin and pseudospin 
which gives rise to rich physics in this unique two dimension system, 
e.g., unusual quantum Hall effect.\cite{zhang_experimental_2005}
Breaking these symmetries could lead to interesting physical phenomena, 
\cite{sodemann_broken_2014,zhou_substrate_2007,young_spin_2012,yao_valley_2008} 
such as gap opening due to pseudospin symmetry breaking in supported graphene,\cite{zhou_substrate_2007}
additional Hall plateau induced by SU(4) spin-pseudospin symmetry breaking due to electron-electron interaction,\cite{young_spin_2012} 
and valley polarization caused by inversion symmetry breaking.\cite{yao_valley_2008}

Graphene can be patterned into a variety of one dimensional nanoribbons, 
of which electronic properties are strongly dependent on the ribbon width and edge patterns.
\cite{PhysRevLett.98.206805,hod_2007,wassmann_2008,lee_2009,li_2008,li_preserving_2013}
Among them, zigzag graphene nanoribbons (ZGNRs) have attracted particular attention owing to their unique edge magnetism,
\cite{son_2006,APL_Duan_2007,PRB_Duan_2008,lee_2009,li_2008,pan_zigzag_2011,li_preserving_2013,jung_2009,magda_2014}
which can be strongly suppressed by metal substrates.
\cite{chen_2012,li_absence_2013,li_electronic_2013,zhang_2013}
Recently, H-terminated ZGNRs on graphene substrate were obtained
by cutting the top layer of graphene bilayer using hydrogen etching method.\cite{li_direct_2014}
Because of the van der Waals interaction between the GNRs and the substrate, 
this type of system provides an ideal platform for studying electronic properties of GNRs.
For instance, edge states were explicitly observed in STM/STS experiments and
a critical width of about 3 nm was revealed 
for the onset of electron-electron correlation between the edges of anti-ferromagnetic (AFM) ZGNRs.\cite{li_direct_2014}

The edge states of ZGNRs also carry unique spin and pseudospin information:
spins are ferromagnetically coupled along each edge but antiferromagnetically coupled between edges,
i.e., along each edge the spin polarized state is spatially located on only one sublattice (pseudospin).
Here we demonstrate, based on density functional theory (DFT) and tight binding (TB) calculations, 
that the unique combination of spin and pseudospin in ZGNRs can also be used to
manipulate properties of graphene.
A half-metallic state is surprisingly obtained in AB (Bernal) stacking of ZGNR on graphene  
as a result of spin- and pseudospin-symmetry breaking caused by the interaction between the GNR and
graphene substrate. 
Remarkably, half-metallic Dirac cones are formed at K (K') near the Fermi level.  

\section{COMPUTATIONAL DETAILS}
To study the electronic and magnetic properties of AFM ZGNRs supported by graphene,
we carried out spin-polarized first-principles
electronic-structure calculations using the Vienna Ab initio Simulation Package.
\cite{kresse_cms_1996,kresse_prb_1996}
The exchange-correlation functional was parametrized in terms of the
local density approximation according to Ceperley and Alder,\cite{ceperley_ground_1980}
and pseudopotentials were constructed by the projector augmented wave method.\cite{bloechl_projector_1994,kresse_ultrasoft_1999}
The one-dimensional Brillouin zone was sampled by a 1$\times$24$\times$1 Monkhorst-Pack mesh for the self-consistent calculations.

\section{Results}
\subsection{DFT calculations}
\begin{figure}
  \includegraphics[width=0.48\textwidth]{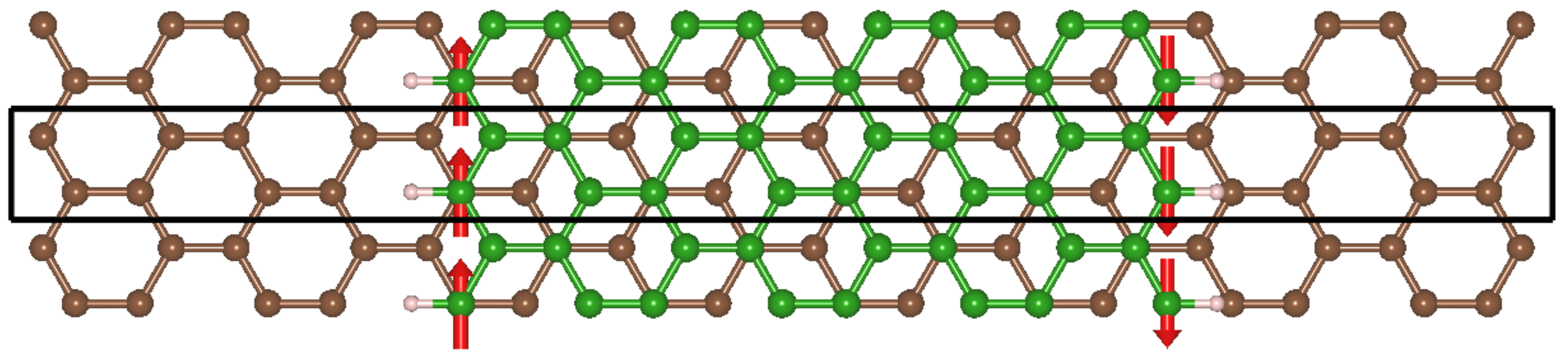}
  \caption{Top view of AB stacking hydrogen-terminated AFM-ZGNR/graphene.
  Green and brown balls represent carbon atoms in the nanoribbon and graphene, respectively.
  Red arrows denote spins on the edge atoms.
  The black box represents the unit cell for calculations. 
  A ZGNR with $n$ zigzag chains is denoted by $n$-ZGNR. The nanoribbon here is 8-ZGNR. 
  }
 \label{fig1}
\end{figure}

Figure~\ref{fig1} schematically illustrates the structure of H-terminated AFM-ZGNR/graphene with AB stacking.
{\color{blue} The ribbon is separated from its periodic images by $\sim$12\ \AA vacuum regions.} 
In both the DFT and the TB calculations
the spin up (down) density corresponds to the majority spin of the left (right) edge of the nanoribbon, 
which is over the top (hollow) sites of the graphene substrate.
Our DFT calculations indicate that  
edge magnetism of AFM-ZGNRs is preserved in the presence of graphene, with a
magnetic moment $\sim$0.13 $\mu_B$ per edge atom for both the freestanding and supported ZGNRs.
Although the magnitudes for the two edge are slightly different for the supported system, 
the difference is small; for example, for 32-ZGNR/graphene, 
the two edges have moments of 0.128 and 0.136 $\mu_B$, respectively.
(The magnetic moments on the edge atoms obtained in the present study are much smaller 
than the one (0.43 $\mu_B$) reported by Ref.\ \onlinecite{son_2006}, 
but consistent with Ref.\ \onlinecite{PRB_Duan_2008}.
However, we note that larger moments ($\sim$0.4 $\mu_B$) can be obtained if the core correction to the
density is neglected.)

\begin{figure}
  \includegraphics[width=0.40\textwidth]{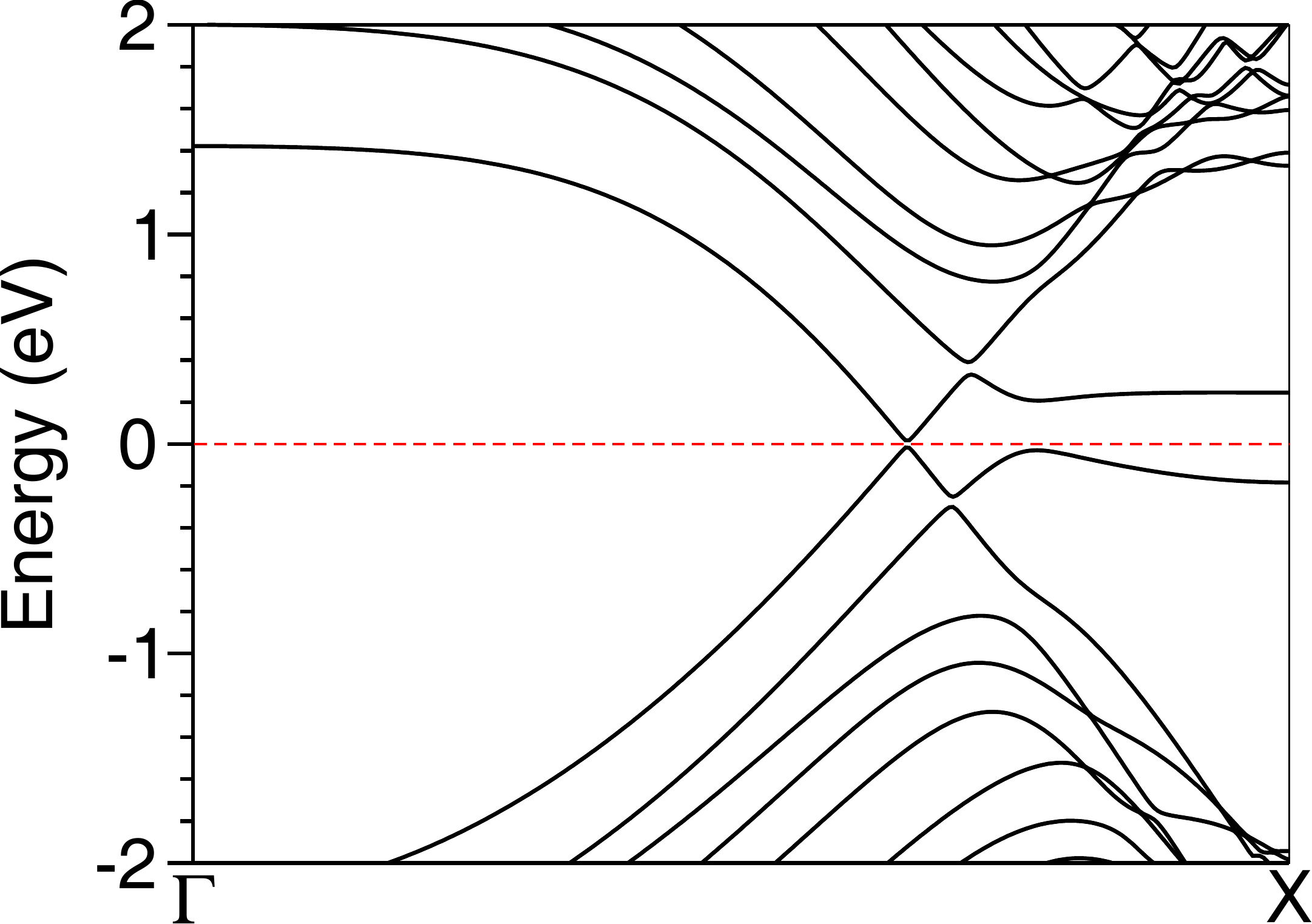}
  \caption{DFT-derived electronic bands for AA stacking of 8-ZGNR/graphene. The Fermi level is represented by a dashed line.
  }
 \label{fig2}
\end{figure}

To see the effects of stacking on the electronic structure of ZGNRs on graphene,  
calculations were also performed for AA stacking 8-ZGNR/graphene. As 
shown in Fig.~\ref{fig2},
AA stacking induces a small gap of about 0.03 eV at the K point. 
The two spins are degenerate since AA stacking maintains the inversion symmetry of the nanoribbon. 

\begin{figure*}
  \includegraphics[width=0.75\textwidth]{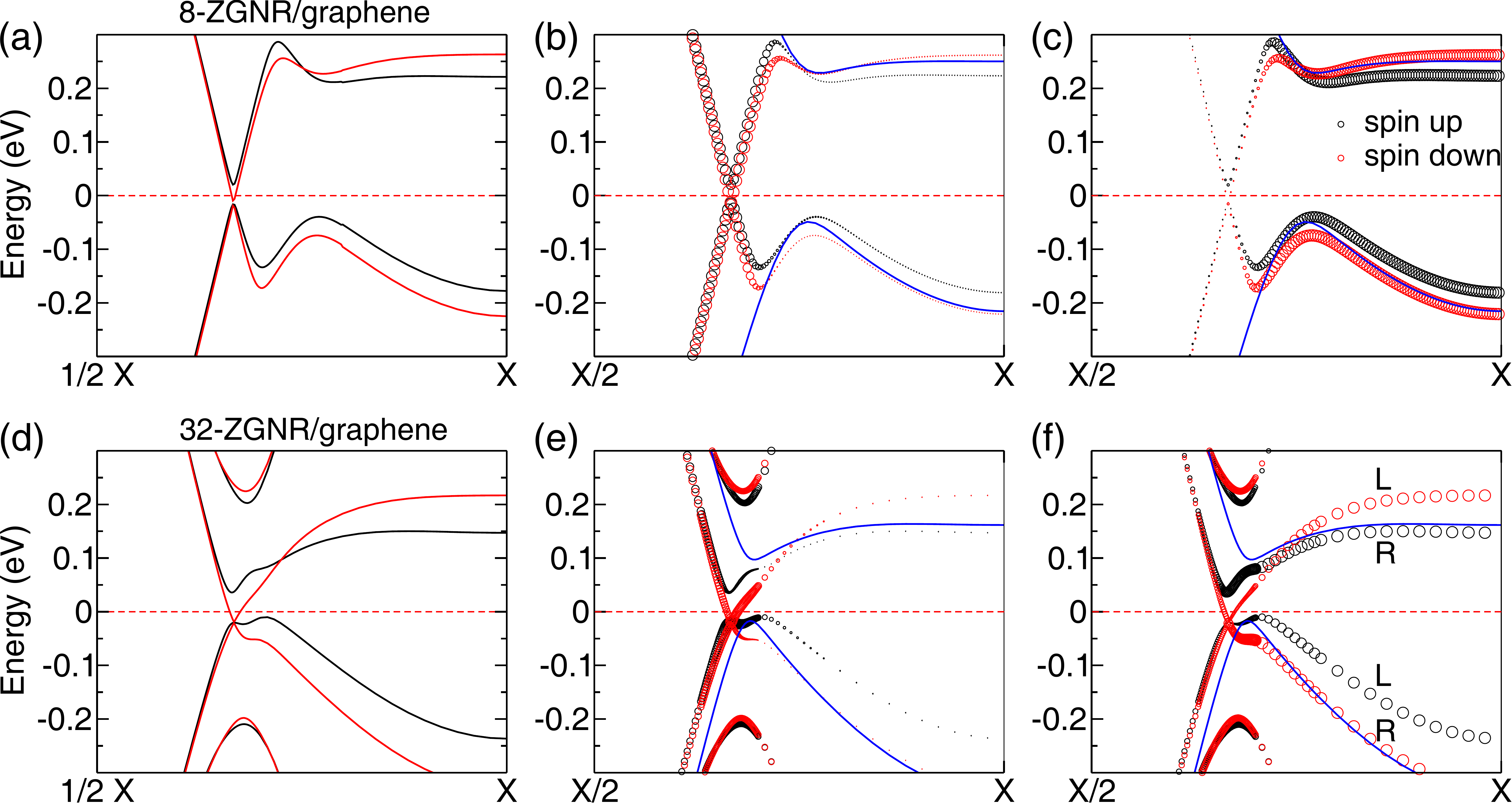}
  \caption{DFT bands for AB stacked AFM-ZGNR/graphene for
  (a) for 8-ZGNR/graphene, and the corresponding (b) graphene- and (c) nanoribbon-weighted bands.
  (d)-(f) Corresponding bands for 32-ZGNR/graphene.
  Bands for the freestanding AFM ZGNR are overlaid as blue solid lines for comparison in (b), (c), (e), and (f).
  Letters L and R stand for the left side and the right side of the nanoribbon, respectively.
  }
 \label{fig3}
\end{figure*}
Figures~\ref{fig3}a and d display the electronic bands for 8- and 32-ZGNR/graphene, respectively.
Compared to AA stacking, a spin splitting is obtained for AB stacking ZGNR/graphene.
The gap opening that occurs at the projection of the K point ($2\pi/3a$) of graphene is spin-dependent:    
it is much larger for spin-up -- the majority spin of the left edge (on top sites) --
than that for spin down, the majority spin of the right edge (in hollow sites).
In particular, for spin down the gap between the valence band and the conduction band is negligibly small at K.
The trend in the gap opening is in fact associated with the location of the edge atom (top sites versus hollow sites).
%(Interchanging the spins on the two edges gives rise to negligible gap for spin up.)
%Remarkably, instead of staying the middle of the gap, 
The Fermi level crosses bands of only one spin state, making the whole system half-metallic as a result of
the nanoribbon-graphene interaction that induces small magnetization in the graphene 
and a net magnetic moment to the whole system.

Figures~\ref{fig3}b, c show the electronic bands for 8-ZGNR/graphene weighted by the localization on the
graphene substrate and the nanoribbon, respectively;
Figs.~\ref{fig3}e and f are the corresponding plots for 32-ZGNR/graphene.
One can see that the bands for 8-ZGNR/graphene at K near the Fermi level are basically graphene bands.
For 32-ZGNR/graphene, both constituents have considerable contributions, as
expected since the bands evolve toward the properties of graphene bilayer as the ribbon size increases. 
Band analysis further reveals that the four bands (including spin) at K are contributed by the hollow sites only, 
consistent with the limit of graphene bilayer.
However, instead of the four-fold degeneracy at the K point in a graphene bilayer, 
for AFM-ZGNRs/graphene the four bands split as a result of spin and pseudospin symmetry breaking.
From Figs.~\ref{fig3}c and f one can see a shift in the edge states near the X point, that is,
the bands for the left edge (top sites) are shifted upward relative to those for the right edge (hollow sites).
The underlying physics is that the AB stacking of AFM-ZGNRs and graphene 
gives rise to different electrostatic potentials for different edges:
Because the edge states are mainly localized to one sublattice, 
especially near the X point, they are either at the top or hollow sites. 
The AB stacking raises the electrostatic potential of the top sites more 
than the hollow sites, resulting in a relative shift in the bands for the two edges.
This mechanism is similar to the case of freestanding AFM ZGNRs under external electric field.\cite{son_2006}
Dramatic changes can be seen as the $k$ point approaches K,  
where the shift of the conduction bands of the two spins is opposite to the case near the X point.
This behavior is related to the interaction between states of AFM ZGNRs and graphene involving 
special spin- and pseudospin-symmetry breaking (see discussions below). 

\begin{figure*}
  \includegraphics[width=0.75\textwidth]{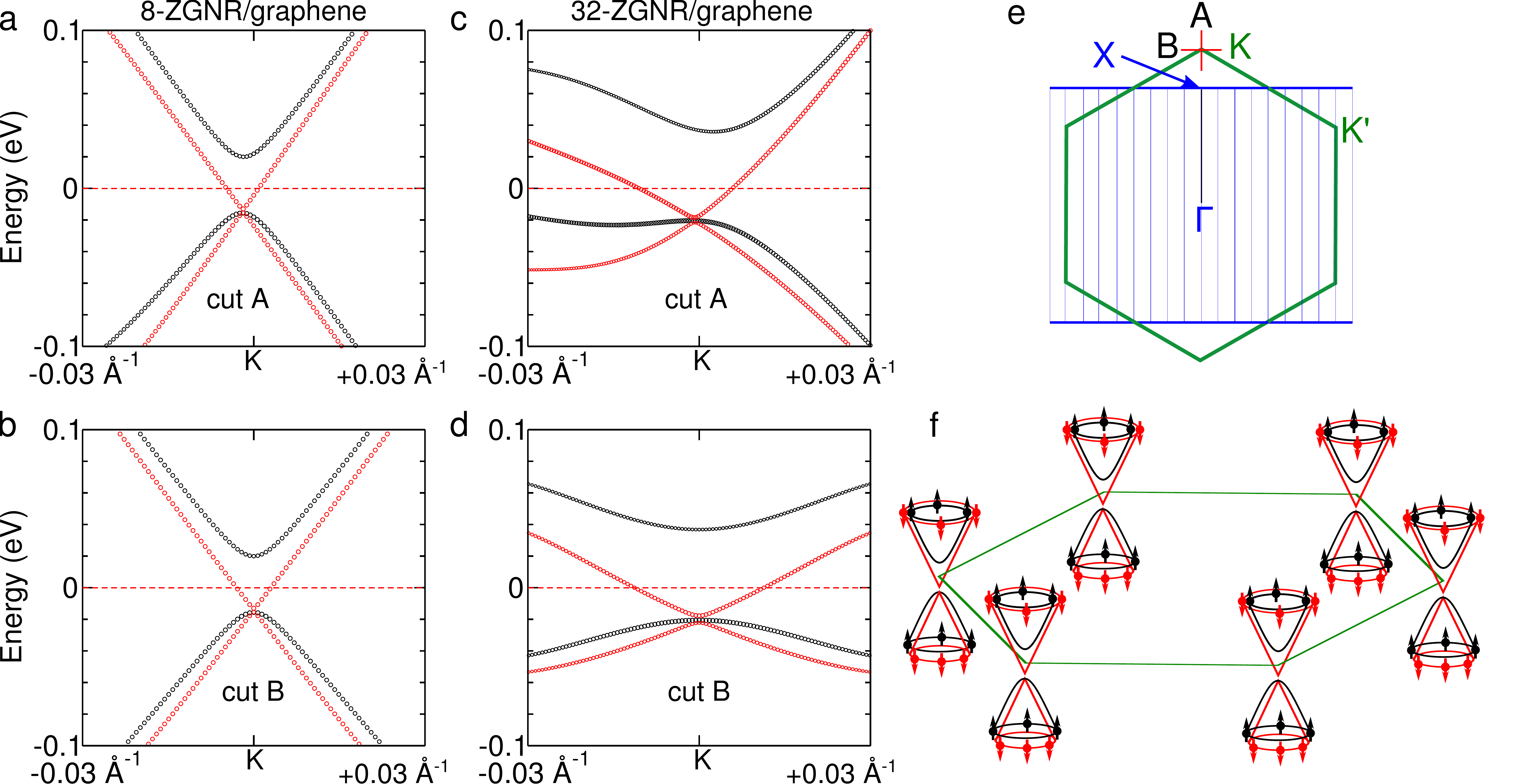}
  \caption{Electronic bands around K.
  (a)$k$-projected bands along $\Gamma$-K (cut A in (e)) and 
  (b) perpendicular to $\Gamma$-K (cut B in (e)) for 8-ZGNR/graphene. 
  (c) and (d) corresponding $k$-projected bands for 32-ZGNR/graphene.
  (e) BZs of 8-ZGNR/graphene and  1$\times$1 graphene. 
  High symmetry points in each cell are also shown. 
  A and B indicate different cuts about K.
  (f) schematic illustration of half-metallic Dirac cone in the BZ of graphene. 
  }
 \label{fig4}
\end{figure*}

To further explore the properties of the bands at K (K'),
Fig.~\ref{fig4} depicts the $k$-projected bands around K for both 8-ZGNR/graphene and 32-ZGNR/graphene,
obtained by projecting the wave functions of AFM-ZGNR/graphene in the rectangular supercell  
onto the 1$\times$1 cell of graphene\cite{bufferlayer,chen_revealing_2014} (c.f., Fig.~\ref{fig4}e).
Half-metallic Dirac states about K exist for all the studied systems. 
However, unlike for ideal graphene, the Dirac cone for ZGNRs/graphene is anisotropic, 
i.e., the linear dispersions along $\Gamma$-K and its perpendicular direction are different,
leading to anisotropic Fermi velocities in ZGNRs/graphene
due to the one-dimensional nature of the nanoribbon.  
Moreover, the linear dispersions for 32-ZGNR/graphene experience a significant renormalization,
as expected since the bands about K approach those of a graphene bilayer.  
(Calculations were also performed about K', which give rise to the same band structures as required by
symmetry and are not shown here.)
Fig.~\ref{fig4}f schematically show the half-metallic Dirac cones at K and K' in the Brillouin zone (BZ) of graphene.

\begin{figure}
  \includegraphics[width=0.40\textwidth]{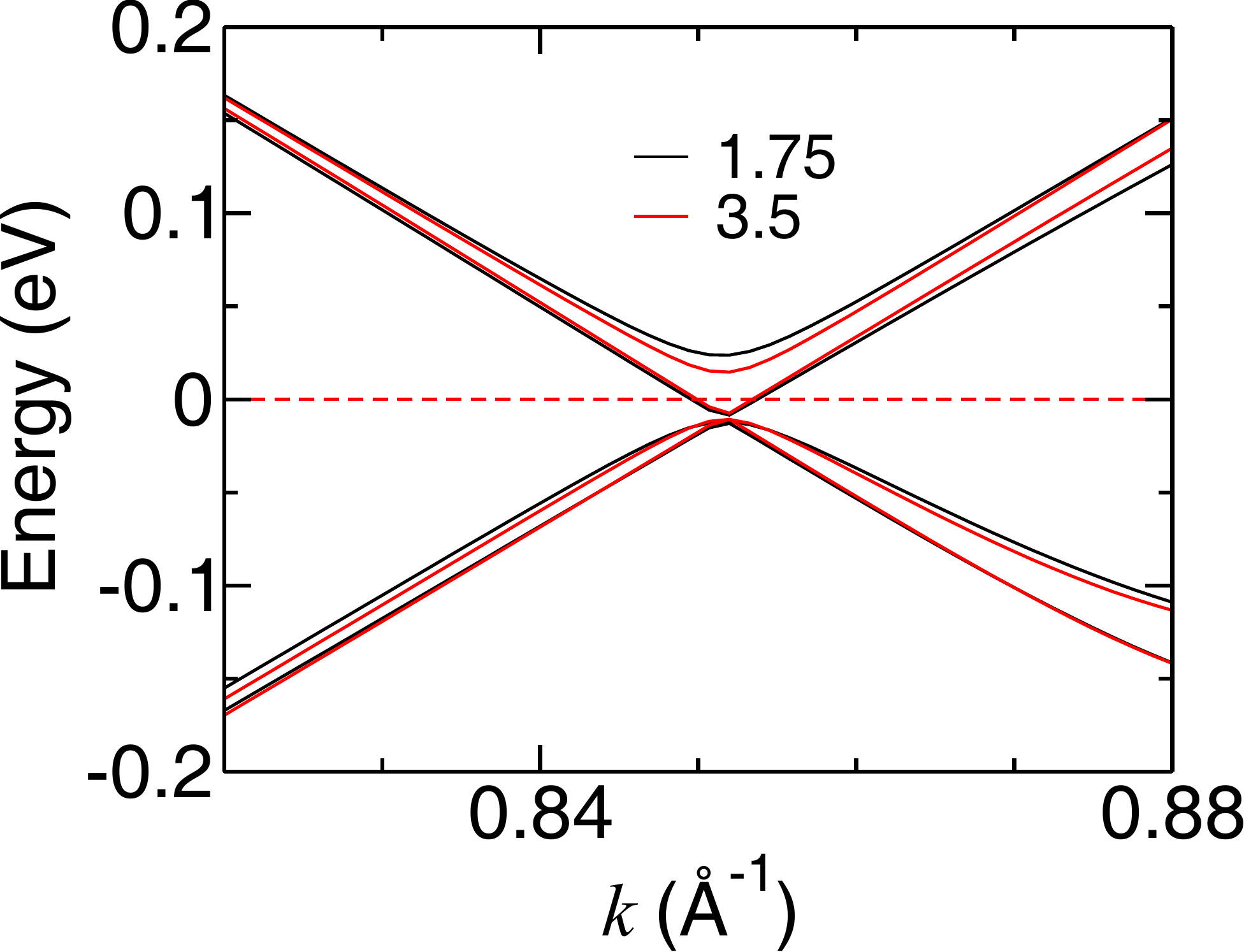}
  \caption{
  Electronic bands along $\Gamma$-X near the point where the half-metallic behavior occurs
  for an 8-AFM-ZGNR on graphene with different supercells corresponding the ratio (1.75 and 3.5) of the
  the area of the graphene substrate to that of the nanoribbon.}
 \label{fig5}
\end{figure}

A question arises whether the half-metallic behavior seen near K is an artifact of the choice of supercell, in
particular the ratio of the graphene substrate to the nanoribbon.
Although a calculation for an AFM-ZGNR on infinite graphene is obviously infeasible,
we performed a calculation for 8-AFM-ZGNR on graphene by doubling the ratio, i.e.,
the ratio of the area of the graphene to that of the nanoribbon is increased to 3.5 from 1.75.
Such a structure gives rise to $\sim$45\ \AA separations between the nanoribbon and its periodic images on
the graphene substrate.   
Figure~\ref{fig5} shows that there are only minor changes in the band structure near K when the ratio is
doubled; in particular, the half-metallic behavior is well preserved.
The gap for spin down, however, is slightly reduced,
as the substrate bands approach those of a single graphene layer in the limit of infinite graphene.
Nevertheless, one may still expect half-metallic behavior as long as there exists
the spin- and pseudospin symmetry breaking which are caused by the presence of the AFM ZGNR and the AB stacking. 

Another possible issue is the effect of vdW dispersion forces between the nanoribbon and the substrate.
To investigate such effects vdW-DF calculations were performed for 8-AFM-ZGNR/graphene
using the method developed by Klime\v{s} and Michaelides,\cite{PhysRevB.83.195131,klimes_chemical_2010}
for which the optB88-vdW functional was used for the exchange functional.
The equilibrium layer distance derived from these calculation is about 3.46\ \AA, 
close to the 3.35\ \AA used for the above calculations, and thus the 
calculated band structure (not shown), is very similar to that shown in Fig.~\ref{fig3}a, including
half-metallic nature.  
 
\subsection{Tight-binding calculations}
The half-metallic states near K originate from the spin and pseudospin-symmetry breaking 
due to the special stacking of the two constituents,
which can be understood by a tight-binding model.
The Hamiltonian for the graphene can be written as  
\begin{equation}
 H^{G} =-\sum_{ij}(t_{ij}a_{i}^{\dag}a_{j}+h.c.) + \sum_{i}\mu_1  a_{i}^{\dag}a_{i}.
 \label{graphene}
\end{equation}
The edge states of ZGNRs can be described by the Hubbard model within the Hartree-Fock approximation
\cite{wakabayashi_2010,hancock_2010}
\begin{equation}
 \begin{split}
 H_{\sigma}^{R} =-\sum_{ij}(t_{ij}c_{i\sigma}^{\dag}c_{j\sigma}+h.c.) \\
 + U\sum_{i}(n_{i\sigma}<n_{i-\sigma}> - \frac{1}{2} <n_{i\sigma}><n_{i-\sigma}>) \\
 + \sum_{i}\mu_2 c_{i\sigma}^{\dag}c_{i\sigma},
 \end{split}
 \label{ribbon}
\end{equation}
where $c_{i\sigma}^{\dag}$ and $c_{i\sigma}$ are creation and annihilation operators for spin $\sigma$ at
site $i$, respectively,
$n_{i\sigma} =c_{i\sigma}^{\dag}c_{i\sigma}$, and
$t_{ij}$ and $U$ denote hopping integrals and on-site Coulomb interaction between electrons, respectively.
The respective chemical potentials are  $\mu_{1,2}$.
The self-consistent solution to Eq.~(\ref{ribbon}) gives rise to 
\begin{equation}
<n_{A\sigma}> = <n_{B-\sigma}> 
\label{den_symm}
\end{equation}
with respect to the symmetry center, 
where $A$ and $B$ are the two sublattices corresponding to the different edges.
Note that Eq.~(\ref{den_symm}) implies spin degeneracy in freestanding AFM ZGNRs.
In our calculations $U$ was set to 2.0 eV and the
$t_{ij}$ include up to the second-nearest-neighbor interaction according to Ref~[\onlinecite{hancock_2010}],
such that electronic bands derived from TB calculations are in good agreement with {\it ab initio} results for the free standing nanoribbons.

The Hamiltonian describing the interaction between the ZGNR and graphene has the form 
\begin{equation}
 H^{int} =-\sum_{ij}(\tau_{ij}a_{i}^{\dag}c_{j}+h.c.),
\end{equation}
where $a_{i}^{\dag}$ and $c_{j}$ are creation and annihilation operators for the two constituents, respectively.
$\tau_{ij}$ describes the hopping between atom $i$ in the nanoribbon and atom $j$ in graphene.
In AB stacking, top-site carbon atoms experience stronger perturbation than those in hollow sites.
This physics can be captured by considering hopping only between the nearest top-top sites,
properly chosen so that TB calculations reproduce the bands of graphene bilayer about the K point.
As a consequence of $H^{int}$ breaking the spin symmetry in Eq.(\ref{den_symm}),
the two edge states of the nanoribbon, which carry different spins, are no long
degenerate.
Similarly, the interaction breaks the spin and pseudospin-symmetry in graphene in the same manner.

\begin{figure}
  \includegraphics[width=0.45\textwidth]{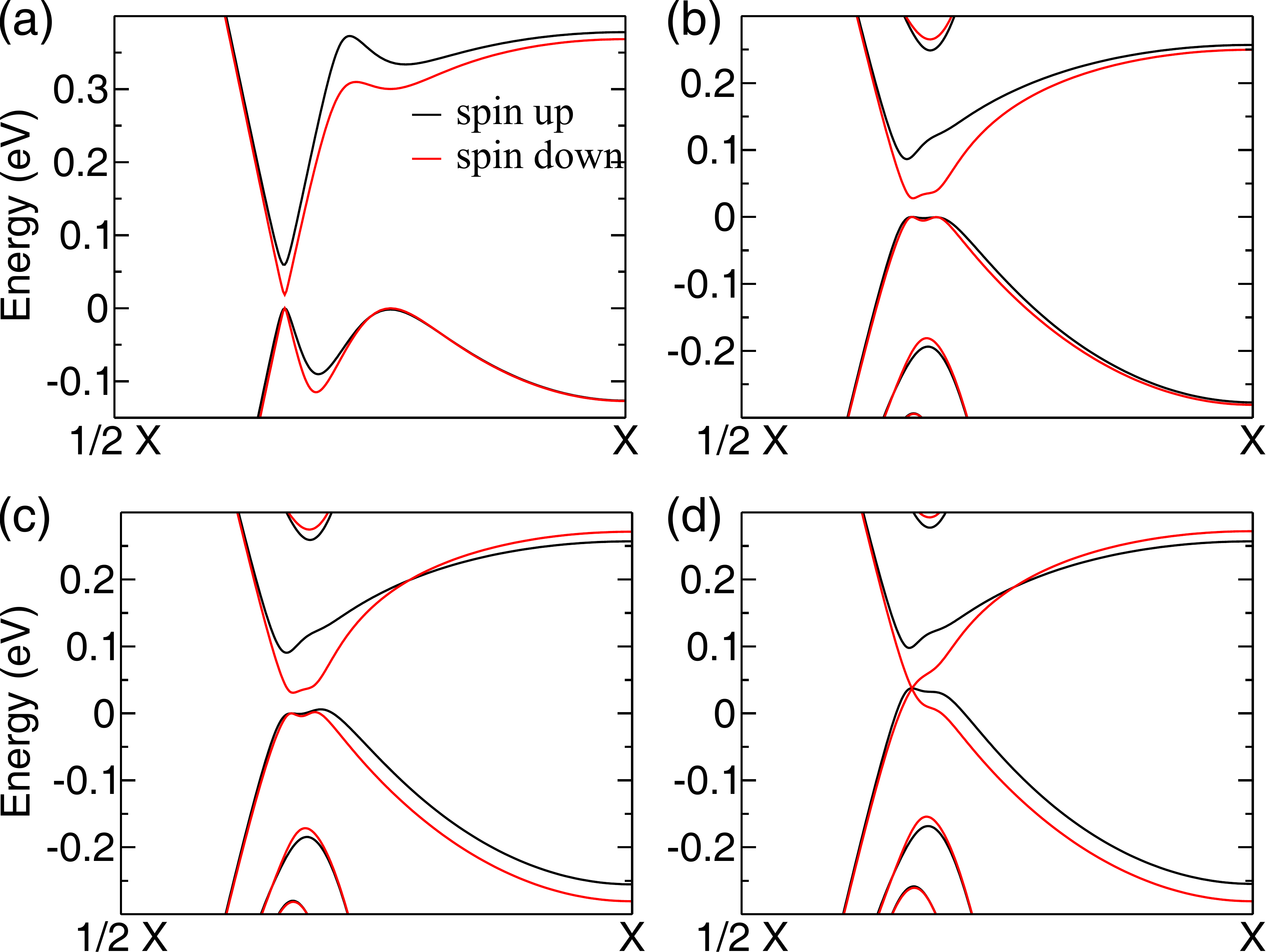}
  \caption{
 TB calculations of AFM-ZGNRs/graphene for (a) 8-ZGNR and (b) 32-ZGNR. 
 (c) Band structure for 32-ZGNR/graphene when the on-site energies for the top and hollow sites are treated differently, and 
 (d) when weak magnetism is induced in the substrate.
  }
 \label{fig6}
\end{figure}
Figures~\ref{fig6}a, b display the electronic bands derived from the TB calculations for 8-ZGNR/graphene 
and 32-ZGNR/graphene, respectively.
These calculations show the same trend as DFT calculations: the gap opening for the the majority spin of
the top sites
is much larger than the one for the other spin.
If the on-site energies for the top and hollow sites are treated differently, as commonly done for
graphene bilayers, 
a band shifting at the Brillouin zone (BZ) boundaries (the X point) is obtained (Fig.~\ref{fig6}c).
Because of magnetic proximity effect, the coupling of an AFM ZGNR to graphene leads to spin polarization in graphene.
In this sense, introducing weak magnetism in graphene shrinks the gap opening (Fig.~\ref{fig6}d).
In particular, the gap for spin down becomes extremely small, consistent with the DFT calculations.

\subsection{Low energy model calculations}
\begin{figure}
  \includegraphics[width=0.48\textwidth]{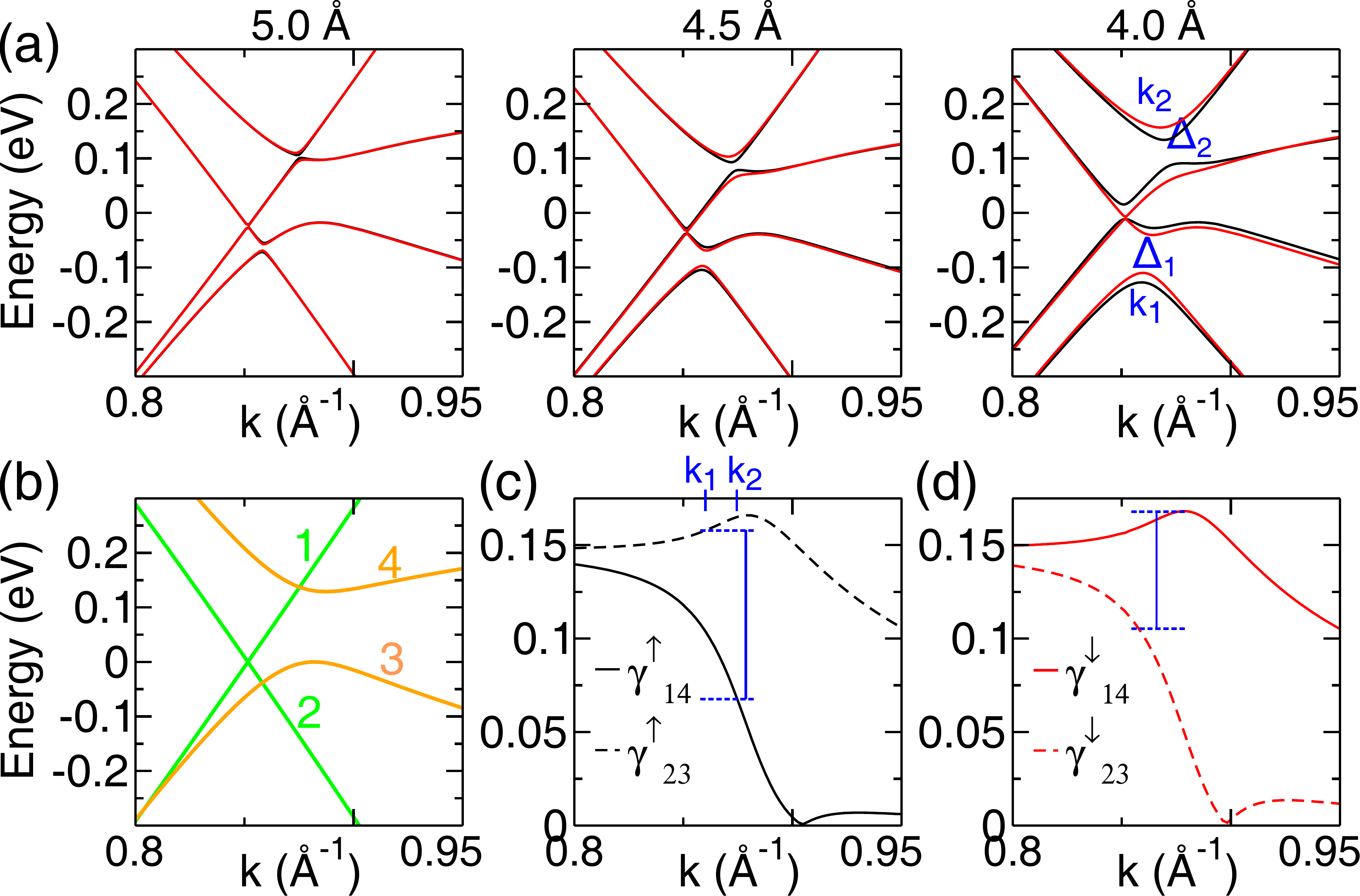}
  \caption{Interaction between bands of AFM-ZGNR and graphene.
(a) DFT-derived band structures for AFM 32-ZGNR/graphene with respect to different interlayer separations
(4.0--5.0 \AA).
Gap openings at $k_1$ and $k_2$ are labeled by $\Delta_1$ and $\Delta_2$, respectively.
(b) TB bands for the isolated systems before interaction.
Bands 1 and 2 are the graphene substrate linear dispersions. 
Bands 3 and 4 are the valence and conduction bands, respectively, of AFM 32-ZGNR.
(c) and (d) $\gamma_{14}$ and $\gamma_{23}$ derived from Eq.~\ref{gamma} for (c) spin up and (d) spin
down. 
The two blue solid lines show the difference between $\gamma_{14}$ at $k_1$ and $\gamma_{23}$ at $k_2$. 
  }
 \label{fig7}
\end{figure}

To see how the nanoribbon-graphene interaction affects the gap openings at K, 
we vary the layer distance gradually from 5 to 4 \AA.
Because the half-metallic feature is more pronounced for larger ZGRNs, 
calculations were carried out for 32-ZGNR/graphene only, with the results are summarized in Fig.~\ref{fig7}a.
The gap openings at K are accompanied by two band splittings at $k_1$ (for the valence band) and $k_2$
(for the conduction band) labeled as $\Delta_1$ and $\Delta_2$, respectively, 
resulting from the interaction between the nanoribbon bands 3 and 4 of Fig.~\ref{fig7}b
and the linear dispersing bands of the graphene substrate.
Based on the bands of the isolated systems shown in Fig.~\ref{fig7}b, 
the band splittings $\Delta_1$ is due to the interaction between bands 2 and 3, 
whereas $\Delta_2$ is attributed to the interaction between bands 1 and 4. 
For spin up, which has a sizable gap at K, $\Delta_1$ is much larger than $\Delta_2$.
For spin down, however, $\Delta_2$ grows and becomes larger than $\Delta_1$ as the interlayer separation
decreases.
Such an enhanced splitting pushes the bonding state (resulting from the interaction between bands 1 and 4) down further, 
reducing the gap with the antibonding state resulting from the interaction of bands 2 and 3.

\begin{figure}
  \includegraphics[width=0.49\textwidth]{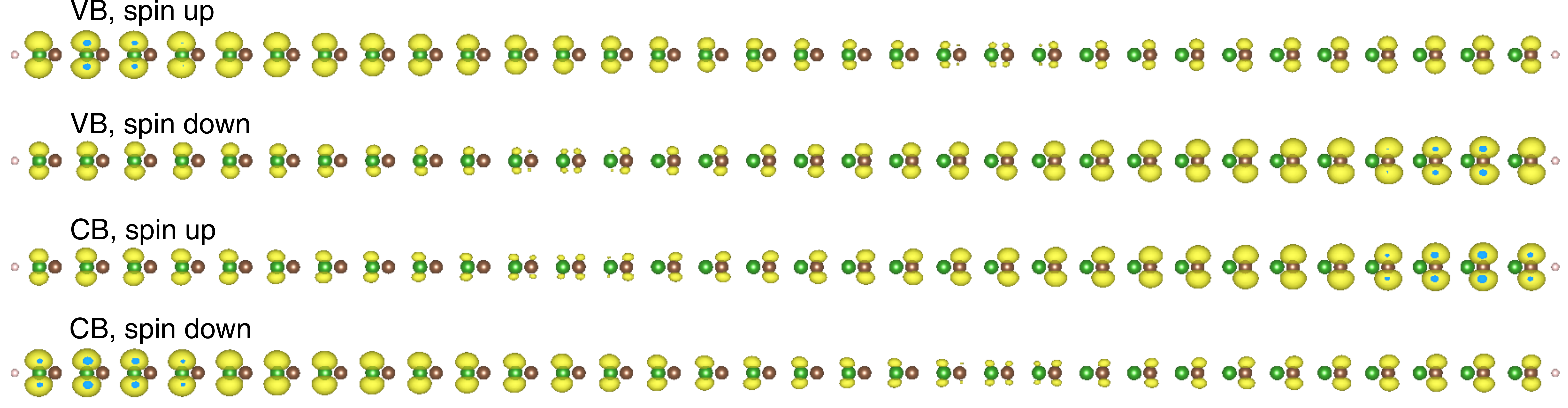}
  \caption{Band-decomposed charge density at k$_2$ for an isolated AFM
32-ZGNR. VB and CB denote the valence band and the conduction band (labeled by 3 and 4, respectively, in Fig.~\ref{fig7}(b).
The two sublattices are colored differently.
  }
 \label{fig8}
\end{figure}

The behavior of $\Delta_1$ and $\Delta_2$ 
is inherently related to the characteristics of the wave functions of the edge states in AFM ZGNR.
Unlike in graphene where the two sublattices make equal contributions to the linear dispersing bands, 
they make asymmetric contributions to the valence and conduction bands of AFM ZGNRs.
At the zone boundary (X point), the valence and conduction band edge states
are completely localized on only one sublattice, either top sites or hollow sites.
As the $k$ point varies along X--K, these states involve an increasing contribution of the other sublattice 
but a large asymmetry in the contributions of the two sublattices remains.
For instance, Fig.~\ref{fig8} shows that the valence band for freestanding 32-ZGNR near the K point  
is mainly contributed by top sites for spin up and by hollow sites for spin down, and
for the conduction band the situation is opposite.
Therefore, one may expect a larger $\Delta_1$ for spin up than spin down, as well as 
a larger $\Delta_2$ for spin down when the coupling to graphene in the AB stacking shown in Fig.~\ref{fig1}.  

To gain further insights into the interaction, 
a 4$\times$4 low energy Hamiltonian for the four bands is constructed. 
\begin{eqnarray}
 H= \left(\begin{array}{cccc}
\epsilon_1     & 0           & H_{13}(k)   & H_{14}(k) \\ 
0              & \epsilon_2  & H_{23}(k)   & H_{24}(k) \\
H_{13}^*(k)    & H_{23}^*(k) & \epsilon_3  & 0         \\
H_{14}^*(k)    & H_{24}^*(k) & 0           & \epsilon_4
\end{array}\right)
\label{low_energy}
\end{eqnarray}
where $H_{ij}(k)$ describe the interaction between the bands of graphene and those of AFM-ZGNRs shown in
Fig.~\ref{fig6}b. The matrix elements
$H_{ij}(k)$ are calculated based on the eigenvectors of the freestanding systems:
\begin{equation}
H_{ij}(k)= <\psi_i^{G}(k)|\hat{H}^{int}|\psi_j^{R}(k)> = \gamma_{ij}(k)e^{i\theta_{ij}(k)}
\label{gamma}
\end{equation}
where $\psi^{G}$ and $\psi^{R}$ are derived from Eq.~(\ref{graphene}) and Eq.~(\ref{ribbon}), respectively,
$\gamma$ and $\theta$ stand for the coupling strength and phase factor, respectively.

The band splittings $\Delta_1$ and $\Delta_2$ are mainly the consequence of $H_{23}$ and $H_{14}$, respectively.
Figures~\ref{fig7}c and d show the magnitudes of $H_{14}$ and $H_{23}$ for the two spins, respectively, i.e., 
$\gamma_{14}^{\sigma}$ and $\gamma_{23}^{\sigma}$.
Opposite trends are observed for the two couplings by a comparison of the two plots:
$\gamma_{23}^{\uparrow}$ is larger than $\gamma_{23}^{\downarrow}$, but
$\gamma_{14}^{\uparrow}$ has a smaller value than $\gamma_{14}^{\downarrow}$.
Thus, one may have a larger $\Delta_1$ for spin up than spin down, 
as well as a larger $\Delta_2$ for spin down than spin up.
Fig.~\ref{fig7}c shows that $\gamma_{23}^{\uparrow}(k_1)$ is larger than $\gamma_{14}^{\uparrow}(k_2)$,
while Fig.~\ref{fig7}d indicates an opposite trend.
Moreover, for spin up the difference between $\gamma_{23}(k_1)$ and $\gamma_{14}(k_2)$ is much larger than 
that for spin down, leading to a noticeable difference between $\Delta_1$ and $\Delta_2$ for spin up,
consistent with the observation in Fig.~\ref{fig7}(a).
The 4$\times$4 Hamiltonian reproduces the TB band structure shown in Fig.~\ref{fig6}.

In summary, half-metallic Dirac cones are found in AB stacking of AFM-ZGNRs/graphene 
by combining tight-binding and DFT calculations. This novel behavior 
results from spin and pseudospin symmetry breaking interactions caused by the particular -- but common --
AB stacking.
The present finding demonstrates that the unique combination of spin and pseudospin in zigzag-graphene-nanoribbons 
can be used to manipulate electronic properties of graphene.
Our results have implications for both fundamental investigations and practical applications; like the
rich physics related to Dirac electrons in graphene, the half-metallic Dirac electrons in AFM-ZGNR/graphene 
may also give rise to extraordinary properties and novel phenomena, such as
 interesting quantum electronic transport found at 
graphene-monolayer/bilayer junctions.\cite{PhysRevB.86.235422}
Similarly, one may expect novel electronic transport properties, with
potential applications in spintronics.  
We anticipate that our results will stimulate further investigations on 
electronic and physical properties of half-metallic Dirac cone in this family. 

\begin{acknowledgments}
This work was supported by the U.S. Department of Energy, Office of Basic Energy Sciences, 
Division of Materials and Engineering under Award DE-FG02-05ER46228. 
\end {acknowledgments}

\bibliography{references}
\bibliographystyle{apsrev4-1}
\end{document}